\documentclass[twocolumn,showpacs,aps]{revtex4}
\usepackage{amsmath,amsthm,amscd}
\usepackage{dcolumn}
\usepackage{graphicx}
\usepackage{epsfig}
\begin{document}
\newcommand{\Qb}{\ensuremath{\mathbf{Q}}}
\newcommand{\qb}{\ensuremath{\mathbf{q}}}
\newcommand{\Sb}{\ensuremath{\mathbf{S}}}
\title{Frustrated classical Heisenberg model in 1 dimension with added nearest-neighbor biquadratic exchange interactions}
\author{T. A. Kaplan}
\affiliation{Department of Physics \& Astronomy and Institute for Quantum Sciences, Michigan State University\\
East Lansing, MI 48824}
\begin{abstract}
The ground state phase diagram is determined for the frustrated classical Heisenberg chain with added
nearest-neighbor biquadratic exchange interactions. There appear ferromagnetic, incommensurate-spiral, and
up-up-down-down phases; a lock-in transition occurs at the spiral boundary. The model contains an isotropic
version of the ANNNI model; it is also closely related to a model proposed for some manganites. The
Luttinger-Tisza method is not obviously useful due to the non-linear weak-constraint problem; however the ground
state is obtained analytically by the exact cluster method of Lyons and Kaplan. The results are compared to the
model of Thorpe and Blume, where the Heisenberg part of the energy is not frustrated.
\end{abstract}
 \maketitle The ANNNI (antiferromagnetic next-nearest-neighbor Ising)
model~\cite{fisher,bak} has Ising spins $S_i^z$, i.e. 2-valued objects, located at points $i$ on a simple cubic
lattice with nearest-neighbor ferromagnetic interactions $J_1$ plus next-nearest-neighbor antiferromagnetic
interactions $J_2$ along one of the cubic directions, say x. Its ground state is the same as that of the linear
chain (translated to all x-chains), whose Hamiltonian is
\begin{equation}
H_{annni}=J_1\sum S_n^z S_{n+1}^z+J_2\sum S_n^z S_{n+2}^z,\label{1}
\end{equation}
$n$ running over the integers, either $-\infty$ to $\infty$ or with periodic boundary conditions; also $J_1<0,
J_2>0$, the latter embodying frustration or competition between the two terms. The ground state phase diagram,
which depends only on $\gamma=J_2/|J_1|$, has a very simple structure: there is ferromagnetic ordering for
$\gamma< 1/2$, up-up-down-down ordering for $\gamma>1/2$.~\cite{fisher,lyons}

The common isotropic version of~(\ref{1}) is the Heisenberg model obtained from~(\ref{1}) by the replacement
$S_n^zS_m^z\rightarrow \Sb_n\cdot\Sb_m$, in the classical version of which the spins are classical unit vectors.
We will consider the classical version, which is in fact the mean field approximation to the quantum
model~\cite{kaplan1}. The corresponding ground state is ferromagnetic for $\gamma<1/4$, spiral for $\gamma >
1/4$, the wave vector $q$ varying continuously from 0 as $\gamma$ increases past 1/4.

Thus, not surprisingly, there is great qualitative difference between the (anisotropic) Ising and (isotropic)
Heisenberg cases. A common way of interpolating between these models is to consider the ``XXZ" Hamiltonian,
obtained from the Heisenberg case by the replacement $J_{nm}\Sb_n\cdot\Sb_m\rightarrow
J_{nm}^x(S_n^xS_m^x+S_n^yS_m^y)+J_{nm}^zS_n^zS_m^z$. This is anisotropic unless $J_{nm}^x=J_{nm}^z$ in general.
I want to consider a different connection, which maintains full isotropy, but nevertheless retains some
characteristics of the Ising case. Namely, add a biquadratic term to the Heisenberg model:
\begin{equation}
H=\sum [J_1\Sb_n\cdot\Sb_{n+1}+J_2\Sb_n\cdot\Sb_{n+2}-a(\Sb_n\cdot\Sb_{n+1})^2],\ \  \Sb_l^2=1.~\label{2}
\end{equation}
This, with $J_1<0, J_2>0$ as above, is the model that will be addressed subsequently. That one may expect
Ising-related ordering for large positive $a$ can be anticipated because for $J_1=J_2=0$, the set of ground
states is the set of collinear states, although there is degeneracy as to which axis all spins are parallel.
Thus the entropy per spin is $\ln 2$ in the thermodynamic limit (the contribution of this rotational degeneracy
disappears in the T.L.), the same as for the (non-interacting) Ising model.

Biquadratic exchange has a long history of being found to be important in certain circumstances. E.g., one of
the earliest works indicating appreciable effect of such interactions is in the paramagnetic resonance
experiments of Harris and Owen~\cite{harris}, that studied the nearest-neighbor-pair spectrum of Mn$^{2+}$ ions
in MgO. They find that a value of $j=0.05J$ in the Hamiltonian $J\Sb_a\cdot\Sb_b-j(\Sb_a\cdot\Sb_b)^2$ gives a
much improved and rather good fit to their measurements. The assumption that the coefficient 0.05 indicates a
small effect would be wrong: In fact the correction to the Heisenberg term is almost a factor of 2 (i.e. 100\%)
for some of the observed and calculated Land\'{e} intervals; this comes from the large spin factors involved.
The microscopic origin and an order-of-magnitude estimate were discussed by Anderson.~\cite{anderson1} For more
recent work see~\cite{bastardis} and references therein, and below. I note, in particular, the consideration by
Thorpe and Blume~\cite{thorpe} of the special case of ~(\ref{2}), $J_2=0$.

The well-known Luttinger-Tisza method appears to be not useful for finding the ground state of~(\ref{2}) because
of the non-linearity introduced into the equations for stationarity of $H$ subject to the weak constraint,
\[\sum_j(J_{ij}-2a_{ij}\Sb_i\cdot\Sb_j)\Sb_j=\lambda\Sb_i.\]
Instead I turn to the rather unknown cluster method of Lyons and Kaplan~\cite{lyons}, which is tractable and
solves the problem rigorously. Briefly recall that method. Assume periodic boundary conditions. Then one easily
verifies that~(\ref{2}) can be rewritten as
\begin{equation}
H=\sum_i h_c(\Sb_{i-1},\Sb_i,\Sb_{i+1}),\label{3}
\end{equation}
where the ``cluster energy"
\begin{eqnarray}
\lefteqn{h_c(\Sb_1,\Sb_2,\Sb_3)=}\nonumber\\
& &\frac{1}{2}\{ J_1(\Sb_1\cdot\Sb_2+\Sb_2\cdot\Sb_3)\nonumber\\
& &-a[(\Sb_1\cdot\Sb_2)^2+(\Sb_2\cdot\Sb_3)^2]\} \label{4} +J_2\Sb_1\cdot\Sb_3\label{4}
\end{eqnarray}
involves 3 neighboring spins. Clearly
\begin{equation}
H\ge\sum_i \min h_c(\Sb_{i-1},\Sb_i,\Sb_{i+1}).\label{5}
\end{equation}
One can easily find the minimum of $h_c$. If the corresponding state ``propagates", i.e. if there is a state of
the whole system such that every set of 3 successive spins gives the minimum $h_c$, then according to~(\ref{5}),
this state will be a ground state of $H$. This is the LK cluster method as applied to the present problem. The
method is not limited to 1 dimension or to periodic Hamiltonians.~\cite{lyons}

Now let's minimize $h_c$. First consider coplanar states, and label the angles made by the end spins with the
central spin $\theta,\theta^\prime$, assumed with no loss of generality to be up, as shown in Fig. 1. The
cluster energy is then
\begin{equation}
h_c(\theta,\theta^\prime)=-\frac{1}{2}(\cos\theta+\cos\theta^\prime)+\gamma\cos(\theta-\theta^\prime)
-\frac{a}{2}(\cos^2\theta+\cos^2\theta^\prime),\label{6}
\end{equation}
where  for simplicity I have put $J_1=-1$ (ferromagnetic), and used the previous definition $\gamma=J_2/|J_1|$.
Differentiating gives the conditions for
\begin{figure}
\includegraphics[height=1.5in]{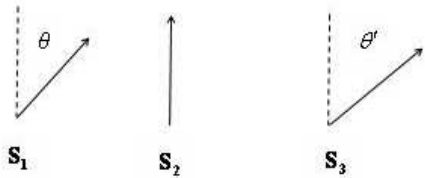}
 \caption{Angles made by the spins in a cluster of 3.}
 \label{fig:spin angles}
\end{figure}
stationarity
\begin{eqnarray}
\frac{1}{2} \sin\theta-\gamma\sin(\theta-\theta^\prime)+a\sin\theta\cos\theta &=& 0\nonumber\\
\frac{1}{2}\sin\theta^\prime+\gamma\sin(\theta-\theta^\prime)+a\sin\theta^\prime\cos\theta^\prime &=&
0.\label{7}
\end{eqnarray}
Solutions are
\begin{eqnarray}
(\theta,\theta^\prime)&=& (0,0),(0,\pi),(\pi,0),(\pi,\pi)\mbox{    (Ising type), and}\nonumber\\
(\theta,\theta^\prime)&=&(\theta_0,-\theta_0)\mbox{        (spiral type), where}\nonumber\\
\cos\theta_0&=&\frac{1}{2(2\gamma-a)}\mbox{    for  } |2(2\gamma-a)|\ge1. \label{8}
\end{eqnarray}

The $(\pi,\pi)$ solution (which leads to the ordinary antiferromagnetic state) is never lowest because we have
assumed $J_1<0$. The (0,0) solution obviously propagates as the ferromagnetic state. The solutions
$(\pi,0),(0,\pi)$, i.e. $(\downarrow,\uparrow,\uparrow),(\uparrow,\uparrow,\downarrow)$ plus their degenerate
reversed spin counterparts can easily be seen to propagate in the up-up-down-down state.~\cite{lyons} The
solution $(\theta_0,-\theta_0)$, degenerate with its uniform rotations, obviously propagates in a simple spiral
\begin{equation}
\Sb_n=\hat{x}\cos n\theta_0+\hat{y}\sin n\theta_0,\label{9}
\end{equation}
$\hat{x},\hat{y}$ being any pair of orthonormal vectors. Such states were first discussed long
ago~\cite{yoshimori,kaplan2,villain}; more generally, for arbitrary Bravais lattices with general $J_{ij}$, it
was shown~\cite{lyons2} that the corresponding spiral, $\hat{x}\cos \qb\cdot\mathbf{n}+\hat{y}\sin
\qb\cdot\mathbf{n}$, minimizes the classical Heisenberg energy for the appropriate wave vector $\qb$.
See~\cite{kaplan1} for a recent review. In the present case, the cluster method provides an alternate proof
(alternative to the Luttinger-Tisza method used in~\cite{lyons2, kaplan1}) for the purely Heisenberg case.
Because of the isotropy of the biquadratic terms, the cluster method accomplishes the proof just as easily.

I list the energies for the various stationary solutions
\begin{eqnarray}
h_{ferro}&=&h_c(0,0)= -1-a+\gamma\nonumber\\
h_{uudd}&=&h_c(0,\pi)=-a-\gamma\nonumber\\
h_{spiral}&=&h_c(\theta_0,-\theta_0)=-\gamma-\frac{1}{4(2\gamma-a)}.\label{10}
\end{eqnarray}
\begin{figure}[h]
\includegraphics[height=2in]{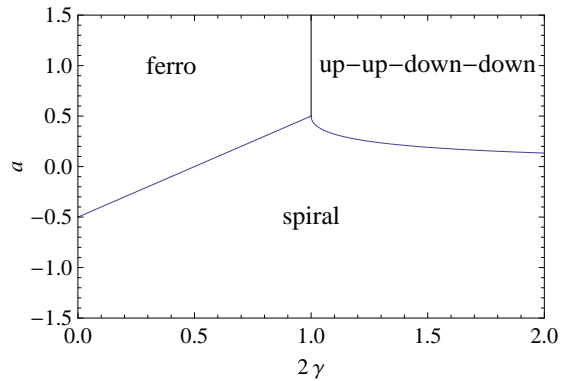}
 \caption{Phase diagram: $a$ vs. $2\gamma$}
 \label{fig:phase diagram}
\end{figure}
The spiral energy holds only for the condition in~(\ref{8}). Equating these energies in pairs yields the
boundaries of the regions shown in~FIG.~\ref{fig:phase diagram}. As a check, to make sure no stationary points
were missed, I calculated the energy difference across boundaries over a mesh of values of $\theta$ and
$\theta^\prime$ varying independently. E.g. I calculated $h_c(\theta,\theta^\prime)-h_{uudd}$ at
$(2\gamma,a)=(1.5, 0.1)$ and (1.5,0.25), the former being a point in the spiral region, the latter in the uudd
region. The former case showed some negative values, the latter only positive values, as must be if the phase
diagram is correct.

\begin{figure}[h]
\includegraphics[height=2in]{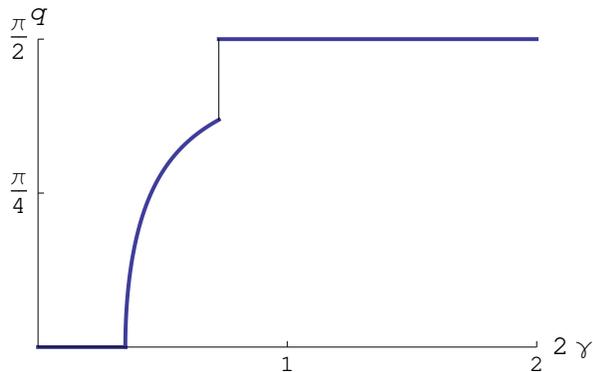}
 \caption{q vs 2$\gamma$ at $a=0.2$; period or wavelength $\equiv 2\pi/q$}
 \label{fig:q vs gamma}
\end{figure}

FIG.~\ref{fig:q vs gamma} shows the variation of $q$ with $\gamma$ for $a=0.2$. In the ferromagnetic and spiral
regions, $q=\theta_0$, the spiral wave vector; in the uudd region, the significance of q is that $2\pi/q$ is the
repeat distance of the spin state. If one moves from inside the spiral region across its boundaries, a lock-in
transition occurs at the boundary, seen in the special case of FIG.~\ref{fig:q vs gamma}. Interest in this
arises because it was thought that such lock-in phenomena were caused by magnetoelastic couplings or anisotropy,
i.e. it results from dependence of spins on the lattice~\cite{gibbs}. While that is probably true in some cases,
the present results indicate another possible cause. Also, as seen in FIG.~\ref{fig:q vs gamma}, the transition
across the spiral-ferro boundary is continuous, while the spiral $\rightarrow$ uudd transition is discontinuous.

In FIG.~\ref{fig:phase diagram}, at $a=0$ it is seen that the ferro $\rightarrow$ spiral transition occurs at
the well-known value $\gamma=1/4$. At $\gamma=0$, the transition ferro $\rightarrow$ spiral occurs at $a=-1/2$,
in agreement with the finding of Thorpe and Blume (TB)~\cite{thorpe} (their $J_2$ is my $-a$); in that work only
$a<0, \gamma=0$ is considered. Furthermore, they find that the state on the line $a < -1/2$ is disordered; this
is not inconsistent with the present finding, which implies only a spiral in the limit $\gamma\rightarrow0$;
\emph{at} $\gamma=0$ the state is indeed highly degenerate, since it depends only on the angle between nearest
neighbors, so that for a given spin $\Sb_n, \Sb_{n+1}$ can lie anywhere on a cone with $\Sb_n$ as axis and
1/2-angle $\theta_0$, giving a (1-dimensionally) macroscopic entropy. Introduction of the 2nd neighbor
Heisenberg interaction removes this degeneracy.\cite{stephenson}

The ferromagnetic transition at $a=-1/2$ on the line $\gamma=0$ shows the following interesting effect: Starting
from $a=0$, adding the extra interaction (the biquadratic terms) of sufficient strength causes  the transition
ferromagnet $\rightarrow$ TB disordered state. This is like the inverse of the ``order-from-disorder"
effect~\cite{villain2}, which is: increasing a commonly thought-to-be disordering parameter, e.g temperature or
impurity concentration, can cause an entropy \emph{reduction}. In the present case, adding the biquadratic terms
increases the entropy as $a$ passes through - 0.5. I.e. the introduction of an additional interaction (usually
thought to remove degeneracy, in the spirit of the Nernst ``theorem"), causes the opposite effect, an
\emph{increase} in entropy: a ``disorder-from-order" effect.

A surprise is that the spiral state continues for negative $\gamma$. The straight-line ferro-spiral boundary,
$a=2\gamma-1/2$, continues to $-\infty$ as $\gamma \rightarrow -\infty$. $q$ or $\theta_0$ vs $\gamma$ at fixed
$a < -1/2$ changes continuously to zero as the ferro-spiral boundary is approached from the right. Nothing
special happens at $\gamma=0$, despite the macroscopic degeneracy at (and only at) that point. The spiral in
this region is caused by the competition between the all-ferromagnetic Heisenberg exchange and the biquadratic
exchange, the latter ``likes" non-collinear spins with angle between nn. spins of $\pi/2$. The nnn. interaction
$\gamma$ removes the macroscopic degeneracy (as for the  antiferromagnetic case).

For large positive $a$ and large $\gamma$ one sees the up-up-down-down phase. This is intuitively reasonable: as
already discussed, the biquadratic terms with $a>0$ are very much like the non-interacting Ising model.

This finding is relevant to the paper, Kimura et al~\cite{kimura}, which studied the frustrated classical
Heisenberg model on a square lattice with nearest-neighbor ferromagnetic interactions $J_1$ and a 2nd-neighbor
antiferromagnetic interaction $J_2$ along one diagonal, (1,1) of the square unit cell. They were seeking the
origin of the ``up-up-down-down" spin state found in certain manganites, in particular HoMnO$_3$. This state
shows spin stripes in the a-b plane lying along (1,-1), varying up,up,down,down as one moves along the (1,1)
direction. They presented a phase diagram that showed this state at $|J_2/J_1|>1$. However it was recently noted
that the correct solution of the assumed model is quite different, the uudd state occurring only in the limit
$|J_2/J_1|\rightarrow\infty$, where it is degenerate with a spiral with a 90$^o$ turn-angle, propagation vector
$\qb$ in the (1,1) direction.~\cite{kaplan3} This realization continued the question as to the source of this
state, and motivated the present study.

In this connection, one should note another path to the uudd state, namely the very different model where the
nearest neighbor exchange varies from ferromagnetic to antiferromagnetic, in continuing periodic fashion. This,
with no other interactions, trivially leads to the uudd state. This one dimensional model is very close to the
mechanism proposed by Zhou and Goodenough~\cite{zhou} for the same manganites discussed in~\cite{kimura}. The
alternating sign of the nearest neighbor exchange interaction in the a-b plane of these materials is argued,
quite reasonably, as being caused by the complex structure of the Jahn-Teller distortion.~\cite{zhou}

I mention two other related works. Girardeau and Popovi\'{c}-Bozi\'{c}~\cite{girardeau} considered the quantum
version of the model of Thorpe and Blume~\cite{thorpe}, showing that in the mean field approximation the
biquadratic terms are not equivalent to replacement by classical spins (unlike the Heisenberg terms). Their
qualitative conclusions are like those of~\cite{thorpe}, particularly with respect to the disordered phase.
Perhaps the earliest paper on the model of combined Heisenberg-biquadratic interactions on a lattice is that due
to Rodbell et. al.~\cite{rodbell} for rock-salt structure antiferromagnets, MnO , NiO. They assumed negative
sign for the biquadratic terms ($a>0$ in my notation), and found large stiffening of the sublattice
magnetization (the stiffening expected for this sign), and within their approximation, good agreement with
experiment. Their model is in a sense closer to the present work in that in these structures there are competing
Heisenberg exchange interactions; however, in these cases the interactions are consistent with collinear spin
states~\cite{smart}, so the qualitative behavior is not similar to that found in the present work.

In summary, the ground state of the frustrated Heisenberg model plus biquadratic exchange interactions on a
linear chain has been solved analytically through an exact cluster method~\cite{lyons}. The case where the
Heisenberg interactions are all ferromagnetic (unfrustrated Heisenberg model) is included. The phase diagram
shows ferromagnetic, spiral, and up-up-down-down spin structures. In the unfrustrated Heisenberg case, the
spiral is caused by the competition between the Heisenberg  and the biquadratic interactions. In this isotropic
model, the periodicity of the spin structures shows a lock-in transformation at the boundary of the spiral
phase.

 The finite temperature behavior of this model, extended to 3 dimensions via the scheme used
in the ANNNI model~\cite{fisher}, (in order to allow long range ordered structures), will probably show
interesting novel behavior. As already noted, the ground state in this higher-dimensional case is solved by the
present results. Also, I expect that extension of the ground-state problem to higher-dimensional simple cubic
models with Heisenberg interactions analogous to those in the model of~\cite{kimura,kaplan3}, and n.n.
biquadratic exchange, should again be tractable via the exact cluster method~\cite{lyons}.

I thank S. D. Bhanu Mahanti and Phil Duxbury for helpful discussions, and Alex Kamenev and Mark Dykman for
encouragement.

\end{document}